\documentclass[11pt]{article}
\usepackage[left=1in,top=1in,right=1in,bottom=1in]{geometry}
\usepackage{times}

\usepackage{url}
\usepackage{verbatim}
\usepackage{amsmath}
\usepackage{amssymb}
\usepackage{amsthm}
\usepackage{rotating}
\usepackage{algorithm}
\usepackage[noend]{algpseudocode}

\usepackage{tabularx}
\usepackage{subfigure}
\usepackage{multirow}
\usepackage{siunitx}

\newtheorem{definition}{Definition}
\newtheorem{theorem}{Theorem}
\newtheorem{proposition}{Proposition}

\begin{document}
\title{Computing Nash Equilibria in Multiplayer DAG-Structured Stochastic Games with Persistent Imperfect Information}
\author{Sam Ganzfried\\
Ganzfried Research\\
sam@ganzfriedresearch.com
}
\date{\vspace{-5ex}}

\maketitle

\begin{abstract}
    Many important real-world settings contain multiple players interacting over an unknown duration with probabilistic state transitions, and are naturally modeled as stochastic games. Prior research on algorithms for stochastic games has focused on two-player zero-sum games, games with perfect information, and games with imperfect-information that is local and does not extend between game states. We present an algorithm for approximating Nash equilibrium in multiplayer general-sum stochastic games with persistent imperfect information that extends throughout game play. We experiment on a 4-player imperfect-information naval strategic planning scenario. Using a new procedure, we are able to demonstrate that our algorithm computes a strategy that closely approximates Nash equilibrium in this game.
\end{abstract}

\section{Introduction}
\label{se:intro}
Many important problems involve multiple agents behaving strategically. In particular, there have been recent applications of game-theoretic algorithms to important problems in national security. These game models and algorithms have differing levels of complexity. Typically these games have two players, and algorithms compute a Stackelberg equilibrium for a model where the ``defender'' acts as the leader and the ``attacker'' as the follower; the goal is to compute an optimal mixed strategy to commit to for the defender, assuming that the attacker will play a best response~\cite{Vorobeychik12:Computing,Vorobeychik14:Computing}. Algorithms have been developed for both zero sum and non-zero-sum games. Computing Stackelberg equilibrium is easier than Nash equilibrium; for example, for two-player normal-form general-sum games, optimal Stackelberg strategies can be computed in polynomial time~\cite{Conitzer06:Computing}, while computing Nash equilibrium is PPAD-hard, and widely conjectured that no polynomial-time algorithms exist~\cite{Chen06:Settling,Daskalakis09:Complexity}. For many realistic scenarios, models are much more complex than even computing Nash equilibrium in two-player general-sum simultaneous games. Many realistic problems in national security involve more than two agents, sequential actions, imperfect information, probabilistic events, and/or repeated interactions of unknown duration.  While some approaches have been developed for models containing a subset of these complexities, we are not aware of any algorithms for solving models that contain all of them. We will present an algorithm for solving a model for a naval strategic planning problem that involves all of these challenges. The model was constructed in consultation with a domain expert, and therefore we have strong reason to believe is realistic. Our model is a 4-player imperfect-information stochastic game. Our model builds on a closely related 4-player perfect-information stochastic game model~\cite{Ganzfried20b:Parallel}. The imperfect information adds a significant additional challenge not modeled by the prior work, which requires development of new algorithms.

The only prior research we are aware of for computing Nash equilibrium in multiplayer imperfect-information stochastic games was approaches that were applied to solve 3-player poker tournaments~\cite{Ganzfried08:Computing,Ganzfried09:Computing}. However, that setting had a property that made it significantly simpler than the general imperfect-information setting. In the poker tournament, the players were dealt a uniform-random poker hand at each game state (which is known privately to only the player), and these private hands did not persist to other states. This allowed them to devise algorithms that operated by essentially solving each state independently without considering effects of the solutions of one state on the private information beliefs at other states (though the algorithms must use value estimations for the transitions to other states). However, these approaches would not work for settings where the private information persists throughout game states, which is often the case in realistic problems. We refer to the case where the imperfect information does not extend beyond the current state a game with \emph{local imperfect information}, and the setting where the imperfect information persists throughout the game---either remaining the same or changing dynamically---as \emph{persistent imperfect information}.

As prior work has shown, the best existing algorithm for multiplayer stochastic games with perfect information and with local imperfect information is a parallel version of an algorithm that combines policy iteration with fictitious play, called \emph{PI-FP}~\cite{Ganzfried09:Computing,Ganzfried20b:Parallel}. This algorithm has been demonstrated to quickly converge to provably very close approximations of Nash equilibrium for a 3-player imperfect-information poker tournament and a 4-player perfect information naval planning problem. PI-FP essentially involves solving each state independently (assuming value estimates for payoffs of the other states) using fictitious play, and then using a procedure analogous to policy iteration (an algorithm for solving single-agent Markov decision processes) to update these value estimates for all states. The parallel version of the algorithm, \emph{Parallel PI-FP}, achieves a linear speedup by doing the state equilibrium computations for multiple states concurrently. While this algorithm has proven to be very successful for these settings, unfortunately it cannot be applied directly to the setting for persistent imperfect information, since it does not account for the private information extending between states.     
   
We observe that our planning problem has the special structure that the game states form a directed acyclic graph (DAG). This allows us to devise a new algorithm that solves each game state sequentially, compute updated type distributions from the state equilibrium strategies at that state, and uses these updated type distributions for solving successive states within the current algorithm iteration. We refer to this new algorithm as \emph{Sequential Topological PI-FP}. Unfortunately this algorithm cannot be parallelized to obtain the linear speedup of Parallel PI-FP; however, we are still able to obtain efficient speed in practice for our problem. 

\section{Imperfect-information naval strategic planning problem}
\label{se:naval}
We will first review the perfect-information naval strategic planning problem that has been previously studied~\cite{Ganzfried20b:Parallel}, and then describe the differences for the imperfect-information version. The game is based on a freedom of navigation scenario in the South China Sea where a set of \emph{blue} players attempts to navigate freely, while a set of \emph{red} players attempt to obstruct this from occurring. 
In our model there is a single blue player and several different red players which have different capabilities (we will specifically focus on the setting where there are three different red players). If a blue player and a subset of the red players happen to navigate to the same location, then a confrontation will ensue, which we call a Hostility Game.

In a Hostility Game, each player can initially select from a number of available actions (which is between 7 and 10 for each player). Certain actions for the blue player are \emph{countered} by certain actions of each of the red players, while others are not.  
Depending on whether the selected actions constitute a counter, there is some probability that the blue player \emph{wins} the confrontation, some probability that the red players win, and some probability that the game repeats. Furthermore, each action of each player has an associated \emph{hostility level}. Initially the game starts in a state of zero hostility, and if it is repeated then the overall hostility level increases by the sum of the hostilities of the selected actions. If the overall hostility level reaches a certain threshold (300), then the game goes into \emph{kinetic mode} and all players achieve a very low payoff (negative 200). If the game ends in a win for the blue player, then the blue player receives a payoff of 100 and the red players receive negative 100 (and vice versa for a red win). Note that the game repeats until either the blue/red players win or the game enters kinetic mode. 
The game model and parameters were constructed from discussions with a domain expert.

\begin{definition}
A perfect-information hostility game is a tuple \\ $G = (N, M, c, b^D, b^U, r^D, r^U, \pi, h, K, \pi^K)$, where
\begin{itemize}
\item $N$ is the set of players. For our initial model we will assume player 1 is a blue player and players 2--4 are red players (P2 is a Warship, P3 is a Security ship, and P4 is an Auxiliary vessel). 
\item $M = \{M_i\}$ is the set of actions, or \emph{moves}, where $M_i$ is the set of moves available to player $i$
\item For $m_i \in M_i$, $c(M_i)$ gives a set of blue moves that are \emph{counter moves} of $m_i$
\item For each blue player move and red player, a probability of blue success/red failure given that the move is defended against (i.e., \emph{countered}), denoted as $b^D$
\item Probability that a move is a blue success/red failure given the move is Undefended against, denoted as $b^U$
\item Probability for a red success/blue failure given the move is defended against, $r^D$
\item Probability for a red success/blue failure given the move is undefended against, $r^U$
\item Real valued payoff for success for each player, $\pi_i$
\item Real-valued hostility level for each move $h(m_i)$
\item Positive real-valued kinetic hostility threshold $K$
\item Real-valued payoffs for each player when game goes into Kinetic mode, $\pi^K_i$
\end{itemize}
\end{definition}

We model hostility game $G$ as a (4-player) stochastic game with a collection of stage games $\{G_n\},$ where $n$ corresponds to the cumulative sum of hostility levels of actions played so far. The game has $K+3$ states: $G_0,\ldots,G_K$, with two additional terminal states B and R for blue and red victories. Depending on whether the blue move is countered, there is a probabilistic outcome for whether the blue player or red player (or neither) will outright win. The game will then transition into terminal states B or R with these probabilities, and then will be over with final payoffs. Otherwise, the game transitions into $G_{n'}$ where $n'$ is the new sum of the hostility levels. If the game reaches $G_K$, the players obtain the kinetic payoff $\pi^K_i$. Thus, the game starts at initial state $G_0$ and after a finite number of time steps will eventually reach one of the terminal states ($B, R, G_K$).

So far, we have described the Perfect-Information Hostility Game (PIHG), which was previously considered. In the real world, often players have some private information that they know but the other players do not. For example, in poker this can be one's private cards, or in an auction one's private valuation for an item. We consider a modification to the PIHG where each player has private information that corresponds to its ``strength,'' or amount of resources it has available. We assume each player has a private \emph{type} $t_i$ from a discrete set $T_i$, where larger values of $t_i$ correspond to increased strength. We assume that the players know only the value of their own type, while each player knows that the other players' private types are drawn from a public distribution. We assume that each player is drawn a private type $t_i$ from the public distribution at the outset of the game, and that this type persists throughout the game's duration. Thus, the game model of the Imperfect-Information Hostility Game (IIHG) is a 4-player imperfect-information stochastic game.

The values of the type parameters affect the probabilities of each player's success during a confrontation, with larger values leading to greater success probabilities. For example, suppose there is an encounter between a blue ship of type $t_b$ and red ship of type $t_r$, and suppose that blue player plays an action $a_b$ that is a counter-move to red's action $a_r$. Then for the PIHG the probability of a blue success would be $p = b^D(a_b)$. In the IIHG we now have that the probability of a blue success will be $p' = p^{\frac{t_r}{t_b}}.$ The other success/failure probabilities are computed analogously. 
Note that for $t_b = t_r$ we have $p' = p$ and the payoffs are the same as for the PIHG. If $t_r > t_b$ then $p' < p$, and similarly if $t_b < t_r$ then $p' > p$. 

\section{Algorithm}
\label{se:algorithm}
We first review the prior algorithm PI-FP, which has been successfully applied to solve PIHG. For standard PI-FP, we first initialize values for each player $i$ at each state $G_j$, $v_i(G_j)$. For example, we can initialize them na\"ively to all be zero or random, or use a more intelligent domain-specific heuristic if one is available. Next, we compute a Nash equilibrium in each game state $G_j$, assuming that the payoff is given by $v_i(G_{j'})$ for the transitions to states $G_{j'} \neq G_j$ (while it is given by the standard game parameters for the terminal payoffs). For Parallel PI-FP these computations are done in parallel, achieving a linear speedup in $k$, the number of available cores. The two main algorithms for approximating Nash equilibrium in multiplayer imperfect-information games are fictitious play~\cite{Brown51:Iterative,Robinson51:Iterative} and counterfactual regret minimization. We opt to use fictitious play due to recent results showing that it leads to significantly closer equilibrium approximation in a variety of multiplayer game classes than CFR~\cite{Ganzfried20:Fictitious}. Note that fictitious play is guaranteed to converge to Nash equilibrium in two-player zero-sum games and certain other game classes, but not in general multiplayer games. While specific counterexamples can be constructed for which it fails to converge in multiplayer games, experiments have shown it to converge consistently in a range of realistic games~\cite{Ganzfried20:Fictitious}. It has also been shown to converge to equilibrium within the context of PI-FP~\cite{Ganzfried09:Computing,Ganzfried20b:Parallel}.

After these game-state equilibrium strategies have been computed using fictitious play, we have now obtained strategies $s_{ij}$ for each player $i$ at each state $G_j$. We next need to update the values $v_i(G_j)$ given these new strategies. This is accomplished using procedures analogous to algorithms for solving Markov decision processes, \emph{policy iteration} and \emph{value iteration}. Value iteration would correspond to performing an essentially local update, where the values are now assigned to the new Nash equilibrium strategy values at that state, which assumed the prior round values for the adjacent states. Policy iteration would involve finding values $v_i(G_j)$ for all players and states that are consistent globally with the new strategies, which can be accomplished by solving a system of equations. These procedures both produce new values for all players at all states, which are then used to re-solve each state $G_j$ again using fictitious play. While various termination criteria can be used for both the inner loop (fictitious play) and outer loop (value updating) portions of the algorithm, we will generally specify a fixed number of iterations in advance to run each, as neither is guaranteed to achieve convergence. When value iteration is used for the inner loop the algorithm is called VI-FP, and for policy iteration PI-FP. For the three-player poker tournament experiments, both VI-FP and PI-FP converged to strategies that constituted an $\epsilon$-equilibrium for very small $\epsilon$~\cite{Ganzfried08:Computing,Ganzfried09:Computing}, with PI-FP converging faster than VI-FP. However for the perfect information hostility game experiments, all versions of VI-FP failed to converge to equilibrium while all versions of PI-FP did~\cite{Ganzfried20b:Parallel}. So for our experiments we will build on PI-FP, which has been demonstrated to have superior performance.

Our main metric for evaluation will be to compute the $\epsilon$, which measures the degree of Nash equilibrium approximation. The value of $\epsilon$ denotes the largest amount that a player can gain by deviating from the strategy profile. In exact Nash equilibrium $\epsilon = 0$, and so naturally our goal is to produce strategies with as small value of $\epsilon$ as possible. We say that the computed strategies constitute an $\epsilon$-equilibrium. Formally, for a given candidate strategy profile $\sigma^*$, define $\epsilon(\sigma^*) = \max_i \max_{\sigma_i \in \Sigma_i} \left[ u_i(\sigma_i,\sigma^*_{-i}) - u_i(\sigma^*_i, \sigma^*_{-i}) \right]$. 	

Now suppose we try to apply standard PI-FP, or Parallel PI-FP, to solve the imperfect-information hostility game (Algorithm~\ref{al:Parallel-PIFP}). Initially we assume that the type probabilities are distributed according to the public prior distribution. For our experiments, we will assume that the prior distribution is uniform over $\{1,\ldots,|T_i|\}$ for each player $i$. We will also assume that the type distribution initially agrees with the prior for all stage games $G_j$, in addition to the initial state $G_0$. We next solve all stage games $G_j$ as before, assuming the types are distributed according to the prior at each game. After solving these games, we must then update the type distributions to make them consistent with the strategies that have just been computed. This can be accomplished by traversing through the states in order $G_0,G_1,\ldots,G_K$ and computing the updated type distribution at each state using Bayes' rule, assuming the type distributions already computed for the preceding states and transitions determined by the computed strategies. After the type distributions are updated at all states to be consistent with the strategies, we can then perform a similar value update step for policy iteration as before by solving the induced system of equations. We continue to repeat these three steps until termination: strategy computation, type distribution update, value update. This modified procedure for solving imperfect-information stochastic games with persistent imperfect-information is given in Algorithm~\ref{al:Parallel-PIFP-II}.  

\begin{algorithm}
\caption{Parallel PI-FP~\cite{Ganzfried20b:Parallel}}
\textbf{Inputs}: Stopping condition $C_S$ for stage game solving, stopping condition $C_V$ for value updating, number of cores $d$
\label{al:Parallel-PIFP}
\begin{algorithmic}
\State $V^0 =$ initializeValues()
\State $i = 0$
\While {$C_V$ not met} 
\State $i = i+1$
\While {$C_S$ not met for each stage game}
\State Run fictitious play on each stage game on $d$ cores (solving $d$ stage games simultaneously) to obtain $S^i$
\EndWhile
\State $V^i = $ evaluateStrategies($S^i$)
\EndWhile
\Return{$S^i$}
\end{algorithmic}
\end{algorithm}  

\begin{algorithm}[!ht]
\caption{Parallel PI-FP for persistent imperfect-information}
\textbf{Inputs}: Stopping condition $C_S$ for stage game solving, stopping condition $C_V$ for value updating, number of cores $d$
\label{al:Parallel-PIFP-II}
\begin{algorithmic}
\State $V^0 =$ initializeValues()
\State $T^0 =$ initializeTypes()
\State $i = 0$
\While {$C_V$ not met} 
\State $i = i+1$
\While {$C_S$ not met for each stage game}
\State Run fictitious play on each stage game with input type distribution according to $T^i$ on $d$ cores (solving $d$ stage games simultaneously) to obtain $S^i$
\EndWhile
\State $T^i =$ updated types for all states consistent with $S^i$
\State $V^i = $ evaluateStrategies($S^i$, $T^i$)
\EndWhile
\Return{$S^i$}
\end{algorithmic}
\end{algorithm} 

Since the type-updating step can be performed efficiently by traversing the game states and applying Bayes' rule, Algorithm~\ref{al:Parallel-PIFP-II} can be performed as efficiently as Algorithm~\ref{al:Parallel-PIFP}, and can obtain the same linear speedup from parallelization. However, there is unfortunately a major problem with Algorithm~\ref{al:Parallel-PIFP-II}. When we are solving a given stage game $G_j$, we are assuming the players have types according to $T_i$, which are consistent with strategies for the states preceding $G_j$ for the strategies under $S_{i-1}$, but not for the strategies that are being newly computed (in parallel) for those states at timestep $i$. So the algorithm would be essentially using stale type distributions that are consistent with old strategies when computing the current strategy at a given state. So the algorithm can still be run, and it may converge; but it will likely obtain poor performance due to the assumption of incorrect type distributions for the game solving.

Fortunately, we can create a new algorithm that correctly accounts for the updated types based on an observation about the structure of the imperfect-information hostility game. In particular, we observe that the game can only transition between states $G_i$ to $G_j$ for $i < j$, since we assume that the hostility levels for all actions are positive and therefore the cumulative hostility can only increase. Thus the states $\{G_i\}$ form a directed acyclic graph (DAG) with a natural topological ordering which is just $G_0,G_1,\ldots,G_K$, where states can only transition to other states that come later in the ordering. This allows us to solve each game state sequentially, compute updated type distributions from the state equilibrium strategies by applying Bayes' rule, and use these updated type distributions for solving successive states within the current algorithm iteration.

Algorithm~\ref{al:ST-PIFP} presents our new algorithm, Sequential Topological PI-FP (ST-PIFP) for solving DAG-structured stochastic games with persistent imperfect information. The algorithm solves each game state sequentially according to the topological ordering, using updated type distributions that have been computed applying Bayes' rule given the strategies and type distributions already computed for the preceding states (for state $G_0$ we assume the original prior type distributions). Then policy iteration is performed to update all game state values globally as before. Note that unfortunately this algorithm is sequential and does not benefit from the linear speedup that the parallel algorithms do. There is an inherent tradeoff in that Algorithm~\ref{al:Parallel-PIFP-II} will run significantly faster, but Algorithm~\ref{al:ST-PIFP} will be significantly more accurate. If Algorithm~\ref{al:ST-PIFP} can be run within the time constraints then it is clearly the preferable choice; however if this is not possible, then Algorithm~\ref{al:Parallel-PIFP-II} can still be run as a last resort. 

\begin{algorithm}[!ht]
\caption{Sequential Topological PI-FP (ST-PIFP)}
\textbf{Inputs}: Stopping condition $C_S$ for stage game solving, stopping condition $C_V$ for value updating
\label{al:ST-PIFP}
\begin{algorithmic}
\State $V^0 =$ initializeValues()
\State $T^0 =$ initializeTypes()
\State $i = 0$
\While {$C_V$ not met} 
\State $i = i+1$
\For {$j = 0$ to $K-1$}
\State $T^i_j =$ updated types for state $G_j$ by applying Bayes' rule assuming the types and strategies for all preceding states. 
\While {$C_S$ not met for stage game $G_j$}
\State Run fictitious play stage game $G_j$ with input type distribution according to $T^i_j$ to obtain $S^i_j$
\EndWhile
\EndFor
\State $V^i = $ evaluateStrategies($S^i$, $T^i$)
\EndWhile
\Return{$S^i$}
\end{algorithmic}
\end{algorithm} 

Note that a DAG-structured stochastic game is different from an \emph{extensive-form game tree}, which is common representation for modeling imperfect-information games. In a game tree each node can have only one parent node, and therefore there can only be at most one path from one node to another. However, this is not the case in a DAG-structured stochastic game; e.g., we can have a path $G_1 \rightarrow G_3 \rightarrow G_6$ as well as just $G_1\rightarrow G_6$, while a tree can only have at most one of those paths. Thus, existing algorithms for solving extensive-form imperfect-information games are not applicable to DAG-structured stochastic games.

So far, we have been assuming that there is a single value for each player for each state. However, in reality the value depends on both the state and on the type distributions of the players. Using the terminology from the partially observable Markov decision processes (POMDP) community, there is a separate value for each \emph{belief state}. While it is intractable to use a model where a separate value is computed for every possible type distribution (there are infinitely many possible distributions for each state), we can still obtain an improvement by associating a separate value depending on each state and assignment of specific types to each player (not distributions of types). It turns out that, assuming the type spaces are small, we can accomplish this with a relatively small modification to Algorithm~\ref{al:ST-PIFP}. Previously, there were $|N|K$ values (one for each player $i$ for each non-terminal state $G_j$), where $|N|$ is the number of players and $K$ is the number of non-terminal states. Assuming that there are $|T_i|$ possible types for each player, our new model will have $K \prod_i |T_i|$ states. If all the $|T_i|$ are equal, then there will be $|N| K |T_i|^{|N|}$ values. While the number of states is exponential in the number of players, we are primarily interested in games with a small number of players. In our experiments we will be using $|T_i| = 2$ and $|N| = 4$ so there would be $64K$ total values.         

Our new algorithm for the setting with type-dependent values in given in Algorithm~\ref{al:ST-PIFP-TDV}. It has a few small differences from Algorithm~\ref{al:ST-PIFP}. First, when running fictitious play to solve the stage games, we now consider the type-dependent game values for transitions to other states. And second, we must now solve a different system of equations for each of the $\prod_i |T_i|$ type combinations. As the bottleneck step of this algorithm is still the game-solving step, this algorithm still takes roughly the same amount of time as Algorithm~\ref{al:ST-PIFP} despite the increased complexity of the value updates.

\begin{algorithm}[!ht]
\caption{ST-PIFP for type-dependent values (ST-PIFP-TDV)}
\textbf{Inputs}: Stopping condition $C_S$ for stage game solving, stopping condition $C_V$ for value updating
\label{al:ST-PIFP-TDV}
\begin{algorithmic}
\State $V^0 =$ initializeValues()
\State $T^0 =$ initializeTypes()
\State $i = 0$
\While {$C_V$ not met} 
\State $i = i+1$
\For {$j = 0$ to $K-1$}
\State $T^i_j =$ updated types for state $G_j$ by applying Bayes' rule assuming the types and strategies for all preceding states. 
\While {$C_S$ not met for stage game $G_j$}
\State Run fictitious play stage game $G_j$ with input type distribution according to $T^i_j$ to obtain $S^i_j$
\EndWhile
\EndFor
\For {Each combination t of types for each player}
\State $V^i_t = $ evaluateStrategies($S^i$, $T^i$)
\EndFor
\EndWhile
\Return{$S^i$}
\end{algorithmic}
\end{algorithm} 

To summarize, for multiplayer stochastic games with perfect information (e.g., the perfect-information hostility game) or with local imperfect information (e.g., the poker tournament), the best approach is Parallel PI-FP. This algorithm is applicable even if the game can have potentially infinite cycles and duration (which was the case for the poker tournament). For multiplayer DAG-structured stochastic games with persistent imperfect information, the best approach is Sequential Topological PI-FP, and in particular the new variant for type-dependent values (ST-PIFP-TDV). We can still apply Parallel PI-FP to these games, but will have worse performance.  

\section{Procedure for computing degree of Nash equilibrium approximation}
\label{se:approximation}
In order to evaluate the strategies computed from our algorithm, we need a procedure to compute the degree of Nash equilibrium approximation, $\epsilon$. For perfect-information stochastic games it turns out that there is a relatively straightforward approach for accomplishing this, based on the observation that the problem of computing a best response for a player is equivalent to solving a Markov decision process (MDP). We can construct and solve a corresponding MDP for each player, and compute the maximum that a player can obtain by deviating from our computed strategies for the initial state $G_0$. This approach is depicted in Algorithm~\ref{al:epc}. It applies a standard version of policy iteration, described in Algorithm~\ref{al:policy-iteration}. Note that while this version of policy iteration is for positive bounded models, we can still apply it straightforwardly to our model which includes negative payoff values, since we do not encounter the problematic situation of potentially infinite cycles of negative rewards. It turns out that Algorithm~\ref{al:epc} can also be applied straightforwardly to stochastic games with local imperfect information. This algorithm was applied to compute the degree of equilibrium approximation in the prior experiments described on the perfect-information hostility game and 3-player imperfect-information poker tournament.         

\begin{algorithm}
\caption{Policy iteration for positive bounded models with expected total-reward criterion~\cite{Puterman05:Markov}}
\label{al:policy-iteration}
\begin{enumerate}
\item Set $n = 0$ and initialize the policy $\pi^0$ so it has nonnegative expected reward.

\item Let $v^n$ be the solution to the system of equations
$$v(i) = r(i) + \sum_j p_{ij}^{\pi^n} v(j)$$
where $p_{ij}^{\pi^n}$ is the probability of moving from state $i$ to state $j$ under policy $\pi^n$.  If there are multiple solutions, let $v^n$ be the minimal nonnegative solution.

\item For each state $s$ with action space $A(s)$, set
$$\pi^{n+1}(s) \in \mathrm{arg}\hspace{-0.1cm}\max_{\hspace{-0.3cm} a \in A(s)} \sum_j p_{ij}^a v^n(j),$$
breaking ties so $\pi^{n+1}(s) = \pi^n(s)$ whenever possible.

\item If $\pi^{n+1}(s) = \pi^n(s)$ for all $s$, stop and set $\pi^* = \pi^n.$  Otherwise increment $n$ by 1 and return to Step~2.
\end{enumerate}
\end{algorithm}

\begin{algorithm}
\caption{\emph{Ex post} check procedure}
\label{al:epc}
\small 
\begin{algorithmic}
\State Create MDP $M$ from the strategy profile $s^*$
\State Run Algorithm~\ref{al:policy-iteration} on $M$ (using initial policy $\pi^0 = s^*$) to get $\pi^*$
\State \Return {$\max_{i \in N} \left[ v_i^{\pi^*_i,s^*_{-i}}(G_0) - v_i^{s^*_i,s^*_{-i}}(G_0) \right]$}
\end{algorithmic}
\end{algorithm}
\normalsize

Unfortunately, Algorithm~\ref{al:epc} can no longer be applied for stochastic games with persistent imperfect information. For these games, computing the best response for each player is equivalent to solving a partially observable Markov decision processes (POMDP), which is significantly more challenging than solving an MDP. It turns out that computing the optimal policy for a finite-horizon POMDP is PSPACE-complete~\cite{Papadimitriou87:Complexity}. The main algorithms are inefficient and typically require an amount of time that is exponential in the problem size~\cite{Cassandra94:Acting}. Common approaches involve transforming the initial POMDP to an MDP with continuous (infinite) state space, where each state of the MDP corresponds to a \emph{belief state} of the POMDP; then this infinite MDP is solved using a version of value iteration where a separate value is associated with each belief state.     

Due to the problem's intractability, we devised a new procedure for our setting that exploits domain-specific information to find optimal policies in the POMDPs which correspond to computing a best response. The algorithm is based on a recursive procedure, presented in Algorithm~\ref{al:epc-pomdp}. The inputs to the procedure are a player $i$, a type $t_i$ for player $i$, a set of type \emph{distributions} $\{\tau_j\}$ for the other players $j \neq i$, the strategies computed by our game-solving algorithm $\{s^*_j\}$, a game state $G_h$, and a time horizon $t$. The procedure outputs the optimal value in the \emph{belief state} for player $i$ when he has type $t_i$ and the opponents have type distribution $\{\tau_j\}$ at hostility state $G_h$ for the POMDP defined by the strategies $\{s^*_j\}$, assuming that a time horizon of $t$ remains. For simplicity of presentation we assume that $T_i =2$ for each player (which is what we will use for our experiments), where $\tau_j$ denotes the probability that player $j$ has type 1, and $1-\tau_j$ the probability of type 2. The algorithm recursively calls itself for updated belief states corresponding to new hostility states that can be transitioned to with horizon $t-1$. As the base case for $t = 0$ we consider only the attainable terminal payoffs from the current state with no additional transitions to new states permitted.    

\begin{algorithm}[!ht]
\caption{ComputeValue($i, t_i, \{\tau_j\}, \{s^*_j\}, G_h, t$)}
\textbf{Inputs}: player $i$, type $t_i$ for player $i$, type distributions for opposing players $\{\tau_j\}$, strategies for opposing players $\{s^*_j\}$, hostility state $G_h$, time horizon $t$ 
\label{al:epc-pomdp}
\begin{algorithmic}
\State max-payoff $= -\infty$
\For {each action $a_i$ for player $i$}
\State payoff $= 0$
\State sum $= 0$
\For {every possible combination of $\alpha_k = \prod_j \gamma_j$, where $\gamma_j \in \{\tau _j, (1-\tau _j)\}$}
\For {every possible terminal outcome o, with payoff $u_i(o)$}
\State payoff += $\alpha_k \cdot u_i(o) \cdot $ probability outcome o is attained when player $i$ takes action $a_i$ and other players follow $\{s^*_j\}$
\State sum is incremented by same excluding $u_i(o)$ factor    
\EndFor
\EndFor
\If {$t \geq 1$}
\For {every possible hostility state $G_{h'} \neq G_h$}
\State $p' = $ total probability we will transition to state $G_{h'}$ when player $i$ takes action $a_i$ and opposing players have type distribution $\{\tau_j\}$ and follow strategies $\{s^*_j\}$
\State $\{\tau'_j\} =$ new type distributions computed using Bayes' rule assuming player $i$ takes action $a_i$ and the game transitions to state $G_{h'}$
\State payoff += $p' \cdot$ ComputeValue($i, t_i, \{\tau'_j\}, \{s^*_j\}, G_{h'}, t-1$)
\State sum += $p'$
\EndFor
\EndIf
\State payoff = payoff / sum
\If {payoff $>$ max-payoff}
\State max-payoff = payoff
\EndIf
\EndFor 
\Return{max-payoff}
\end{algorithmic}
\end{algorithm} 

For the case where the prior type distribution is uniform (all values equal to $\frac{1}{|T_i|}$, which is what we use in our experiments), we apply Algorithm~\ref{al:epc-pomdp} as follows. For each player $i$ and each type $t_i \in T_i$ we apply Algorithm~\ref{al:epc-pomdp}, assuming that each opposing player has type $t_j$ with probability $\tau_j = \frac{1}{|T_j|}$, using the initial game state $G_0$ and time horizon $t$. Call the result $V_{t_i}$. Then the optimal value for player $i$ is $V_i = \frac{\sum_{t_i} V_{t_i}}{|T_i|}$. We repeatedly compute $V_i$ for $t = 0,1,2\ldots$ until it (hopefully) converges. We can then compare the converged values of $V_i$ for each player to the expected payoff for the player under the computed strategy profile, $V^*_i = u_i(s^*)$. We then define $\epsilon_i = V_i - V^*_i$, and $\epsilon = \max _i \epsilon_i$. We were able to apply several implementation enhancements to improve the efficiency of Algorithm~\ref{al:epc-pomdp} for the Imperfect-Information Hostility Game. These included precomputing tables of coefficients for transition and terminal payoff probabilities and type indices, only iterating over future states with $h' > h$, and ignoring states $G_{h'}$ with extremely small transition probability from $G_h$ (e.g., below 0.01).                

\section{Experiments}
\label{se:experiments}
We ran Algorithm~\ref{al:ST-PIFP} (ST-PIFP) and Algorithm~\ref{al:ST-PIFP-TDV} (ST-PIFP-TDV) using 10,000 iterations of fictitious play each for 25 iterations on a single core on a laptop. Each iteration of ST-PIFP took around 20 minutes, while each iteration of ST-PIFP-TDV took around 23 minutes (using $K = 300$). So as predicted, the added complexity of including type-dependent values only led to a small increase in runtime because the running time of the bottleneck game-solving step remained the same. All of our game parameters were the same as for the previously-considered perfect-information version (e.g., 1 blue and 3 red players each with 7--10 actions available, payoffs of +100/-100 for a win/loss and -200 for Kinetic mode, fixed action hostilities and success probabilities which are modified to incorporate the players' types), except we used a smaller version with $K = 150$ to obtain convergence with Algorithm~\ref{al:epc-pomdp}. (For $K = 300$ the algorithm converged for the first few strategy iterations but took too long to converge for later strategy iterations due to the greater degree of randomization in the later strategies, leading them to have weight on a larger number of transition sequences that must be considered.)  We assume each player has $|T_i| = 2$ available types, which are initially uniform random according to the public prior, and we initialize all values for both algorithms to be zero. 

Results for both algorithms after 10 iterations for $K = 150$ are given in Table~\ref{ta:results}. For each algorithm, we report $V^A_i$---the expected payoffs for player $i$ under the computed strategies---as well as $V^O_i$---the optimal payoff according to Algorithm~\ref{al:epc-pomdp}. The numbers in parentheses indicate the values for each type, which are averaged to obtain the value. We see that for ST-PIFP we have $(\epsilon_1,\epsilon_2,\epsilon_3,\epsilon_4) = (4.008, 1.126, 5.472, 0.228)$, giving $\epsilon =\max_i \epsilon_i = 5.472$, while for ST-PIFP-TDV we have $(\epsilon_1,\epsilon_2,\epsilon_3,\epsilon_4) = (0.064, 0.903, 0.538, 0.450)$, giving $\epsilon = \max_i \epsilon_i = 0.903$. Note that the smallest payoff magnitude in the game is 100 and these values correspond to approximately 5\% and 1\% of that quantity respectively. We expect that the $\epsilon$ would quickly converge to 0 with more strategy iterations even for larger games, as in the experiments on PIHG. The results indicate that utilizing type-dependent values does in fact lead to better strategies. As expected we observe that higher type values produce greater expected payoff. 
Interestingly we observe that the overall payoffs of the strategies for all red players are identical (for both algorithms), yet they differ for each type.

\begin{table}[!ht]
\centering
\begin{tabular}{|*{3}{c|}} \hline
 &ST-PIFP &ST-PIFP-TDV \\ \hline
$V^A_1$ &-94.642 (-98.271, -91.034) &-92.840 (-96.103, -89.576) \\ \hline
$V^O_1$ &-90.634 (-97.034, -84.234) &-92.776 (-96.099, -89.454)\\ \hline
$V^A_2$ &4.915 (-23.778, 33.608) &9.681 (-20.949, 40.311) \\ \hline
$V^O_2$ &6.041 (-22.383, 34.465)  &10.584 (19.807, 40.974) \\ \hline
$V^A_3$ &4.915 (-9.961, 19.791) &9.681 (-1.916, 21.278) \\ \hline
$V^O_3$ &10.387 (-3.405, 24.180) &10.219 (-2.035, 22.474)\\ \hline
$V^A_4$ &4.915 (-7.620, 17.450) &9.681 (-2.512, 21.874) \\ \hline
$V^O_4$ &5.143 (-7.529, 17.815)  &10.131 (-1.105, 21.366) \\ \hline
\end{tabular}
\caption{$V^A_i$ is the expected payoff to player $i$ under the strategies computed by the algorithm given in the column, and $V^O_i$ is the optimal payoff to player $i$ according to Algorithm~\ref{al:epc-pomdp}. Numbers in parentheses are the values given types 1 and 2 (which are averaged to produce the indicated value).}
\label{ta:results}
\end{table} 
    
\section{Conclusion}
\label{se:conclusion}
We presented new algorithms for computing Nash equilibrium in DAG-structured stochastic games with persistent imperfect information, which we applied to closely approximate Nash equilibrium strategies in a realistic naval planning problem devised by a domain expert. We evaluated the computed strategies with a new domain-specific procedure for solving the induced POMDP. In the future we would like to create a generalized approach that works for a broader class of games beyond those with a DAG-structured state space---ideally to all stochastic games including those with infinite cycles. We would also like to improve Algorithm~\ref{al:epc-pomdp} to more efficiently solve the induced POMDP for strategy evaluation. We expect our approaches to be broadly applicable to imperfect-information stochastic games beyond the specific game considered. 

\bibliographystyle{plain}
\bibliography{C://FromBackup/Research/refs/dairefs}

\end{document}